# ECG-PPS: Privacy Preserving Disease Diagnosis and Monitoring System for Real-Time ECG Signal

1st Beyazit B Yuksel
*Computer Engineering*
*Istanbul Technical University*
Istanbul, Turkey
yukselbe18@itu.edu.tr

2nd Ayse Yılmazer-Metin
*Computer Engineering*
*Istanbul Technical University*
Istanbul, Turkey
yilmazerayse@itu.edu.tr

*Abstract*— This study introduces the development of a state-of-the-art, real-time ECG monitoring and analysis system, incorporating cutting-edge medical technology and innovative data security measures. Our system performs three distinct functions: real-time ECG monitoring and disease detection, encrypted storage and synchronized visualization, and statistical analysis on encrypted data. At its core, the system uses a three-lead ECG preamplifier connected through a serial port to capture, display, and record real-time ECG data. These signals are securely stored in the cloud using robust encryption methods. Authorized medical personnel can access and decrypt this data on their computers, with AES encryption ensuring synchronized real-time data tracking and visualization. Furthermore, the system performs statistical operations on the ECG data stored in the cloud without decrypting it, using Fully Homomorphic Encryption (FHE). This enables privacy-preserving data analysis while ensuring the security and confidentiality of patient information. By integrating these independent functions, our system significantly enhances the security and efficiency of health monitoring. It supports critical tasks such as disease detection, patient monitoring, and preliminary intervention, all while upholding stringent data privacy standards. We provided detailed discussions on the system's architecture, hardware configuration, software implementation, and clinical performance. The results highlight the potential of this system to improve patient care through secure and efficient ECG monitoring and analysis. This work represents a significant leap forward in medical technology. By incorporating FHE into both data transmission and storage processes, we ensure continuous encryption of data throughout its lifecycle while enabling real-time disease diagnosis. Our entire architecture is available as open-source, encouraging further research and development in this vital field.

## I. Introduction

Monitoring the electrical activity of the heart through electrocardiography (ECG) is essential for diagnosing various heart conditions. With the rise of remote monitoring devices, securing ECG data is critical. Traditional systems often lack robust security, leaving them vulnerable to breaches and unauthorized access. These systems may also struggle with real-time processing, limiting diagnostic effectiveness. Current encryption techniques offer basic security but fail against sophisticated attacks and do not fully protect data privacy during transmission over public networks.

Our study introduces a secure, real-time ECG monitoring system that performs three distinct functions: real-time pulse measurement and heart rhythm detection using a TensorFlow model trained on the MIT-BIH Arrhythmia Database, AES encryption for synchronized real-time data tracking and storage, and Fully Homomorphic Encryption (FHE) for statistical analysis on encrypted data.

First, the system enables real-time pulse measurement and heart rhythm detection. It uses a three-lead ECG preamplifier connected through a serial port to capture, display, and record real-time ECG data. A TensorFlow model trained on the MIT-BIH Arrhythmia Database is employed to detect diseases based on heart rate variability. This allows for accurate, real-time disease detection and patient monitoring. Second, the system uses AES encryption to ensure the secure storage and synchronized visualization of ECG data. Real-time ECG signals are encrypted and stored securely in the cloud. Authorized medical personnel can decrypt and access this data on their computers. AES encryption ensures synchronized real-time data tracking and visualization, maintaining data integrity and confidentiality. Third, the system performs statistical operations on the ECG data stored in the cloud without decrypting it, using Fully Homomorphic Encryption (FHE). This enables privacy-preserving data analysis while ensuring the security and confidentiality of patient information. By allowing computations on encrypted data, FHE ensures that patient data remains secure throughout its lifecycle.

The primary objectives of this study are:

- To enable real-time pulse measurement and heart rhythm detection using a TensorFlow model trained on the MIT-BIH Arrhythmia Database.

- To implement robust AES encryption techniques to secure ECG data, ensuring synchronized real-time data tracking and visualization.

- To perform statistical analyses on data encrypted with FHE, preserving data privacy and security.

By integrating these independent functions, our system significantly enhances the security and efficiency of health monitoring. This methodology facilitates essential functions like disease diagnosis, patient surveillance, and initial intervention, while maintaining rigorous data privacy protocols. Our work offers an open-source framework and operating principles for those wishing to conduct research in the areas of real-time patient monitoring and privacy-preserving disease diagnosis.

The rest of the paper is organized as follows: Section II provides a comprehensive literature review, discussing previous works related to ECG signal processing, encryption techniques, and machine learning models for disease detection. Section III describes the preliminary design and implementation of the proposed real-time ECG monitoring system, including detailed information on the hardware and software components, data acquisition methods, and the implementation of encryption for secure data storage. Section IV presents the experimental results, where the system's performance is evaluated using the MIT-BIH Arrhythmia Database. This includes accuracy, computational efficiency, and the impact of Fully Homomorphic Encryption (FHE) on real-time data processing. Section V concludes the paper with a summary of the key findings and highlights the contributions and limitations of this work. Finally, Section VI suggests potential future directions to further enhance the system's performance and scalability in real-world healthcare applications.

## II. Literature Review

The integration of real-time ECG signal encryption, storage, and analysis using fully homomorphic encryption (FHE) and privacy-preserving methods has a significant focus in recent biomedical research. This literature review summarizes the key advancements and differentiates our approach from existing methodologies.

### A. Fully Homomorphic Encryption in Healthcare

Fully Homomorphic Encryption (FHE) has revolutionized data security, particularly in healthcare, by enabling computations on encrypted data without decrypting it. This preserves patient privacy while allowing for meaningful data analysis and pro-cessing. Foundational work by Craig Gentry demonstrated the feasibility of FHE, leading to its application in various domains, including medical data analysis [3]. A systematic review by Munjal and Bhatia highlights the evolution and categorization of homomorphic encryption techniques, emphasizing their potential in healthcare for secure and efficient data processing. The review discusses various homomorphic encryption schemes, including partial, somewhat, and fully homomorphic encryp-tion, and compares their computational overhead and security features in healthcare applications [4]. Our work builds on these principles, applying FHE to real-time ECG data for secure monitoring and analysis.

### B. Privacy-Preserving Disease Diagnosis

The need for privacy-preserving techniques in disease diagnosis is critical due to the sensitive nature of health data. Various studies have explored methods to ensure data privacy while enabling accurate diagnosis. For instance, Harinee and Mahendran focused on encrypting ECG signals using cellular automata before transmission to prevent unauthorized access during data transfer [5]. This method ensures data in-tegrity and confidentiality, which is vital for remote paitent health monitoring systems. Albalawi et al. investigated privacy-preserving methods applied in electrocardio-gram (ECG) monitoring, focusing on intelligent arrhythmia detection [6]. Kim et al. proposed a new approach that uses differential privacy (DP) as well as homomorphic encryption to improve data security [7]. However, all previous works do not perform real-time processing on encrypted data. In our study, we extend these methodologies by incorporating FHE, allowing direct processing of encrypted ECG data without decryption. This not only protects data privacy but also supports real-time analysis, improving the security and efficiency of health monitoring systems.

### C. ECG Signal Transmission and Storage

Researchers have extensively studied how to securely transmit and store ECG signals. Methods such as AES encryption and differential secrecy used to protect data during transmission and storage. Hameed et al. and Shaikh et al. have demonstrated the effectiveness of these encryption techniques in protecting ECG data against various attacks [8] [9]. Our system leverages these encryption methods but takes them even further, using FHE for both transmission and storage. This ensures that data remains encrypted throughout its lifecycle, providing a higher level of security and enabling secure calculations on encrypted data for real-time analysis and diagnostics.

### D. Secure and Efficient Health Monitoring

The combination of secure data encryption and efficient health monitoring is very important for modern healthcare systems. Studies by Lin et al. and Qin et al. have shown that remote monitoring systems can greatly improve patient care with continuous monitoring and early disease detection [10][11]. Our system focuses on real-time data collection, encryption, and analysis using advanced cryptographic methods. By using Fully Homomorphic Encryption (FHE), our system keeps data private without sacrificing the accuracy of pulse measurement and heart rhythm detection. This method provides a strong framework for secure and efficient health monitoring, which can used in different healthcare environments.

Despite significant advancements in secure ECG data transmission and analysis, our work distinguishes itself by integrating fully homomorphic encryption and AES. This integration offers a comprehensive, privacy-preserving solution for real-time health monitoring and analysis. By ensuring the security of sensitive health data throughout its lifecycle, our approach supports the development of secure and efficient healthcare systems.

## III. Preliminary

This section details the preliminary design and implementation of a real-time ECG monitoring and analysis system. The system, depicted in Fig. 1, comprises a comprehensive software and hardware setup to read ECG signals from a three-lead preamplifier attached to the human body. It acquires data through a serial port and displays it graphically in real-time on a user interface. After obtaining the real-time ECG signal, it can perform 3 different functions.

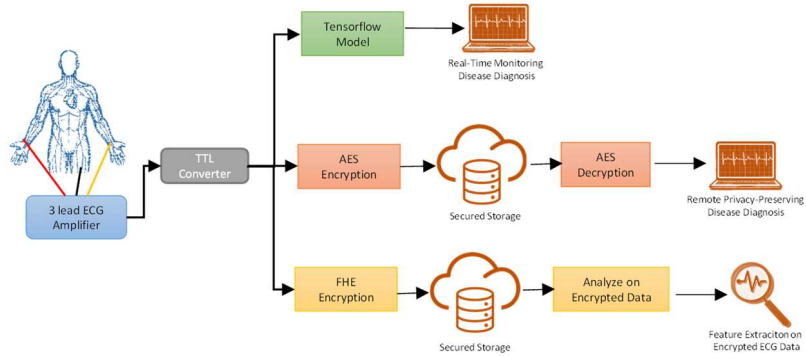

Fig. 1.    System Overview

### A. Hardware Design

The core component of the system design is the ECG preamplifier, which captures the heart's electrical activity using a three-lead system. The leads, connected using a method called Einthoven's triangle and placed on the patient's body, detect voltage differences resulting from the heart's activity. These captured signals are transmitted to the computer via a serial port connection. This preamplifier enhances the weak electrical signals produced by the heart, making them suitable for further processing. A three-lead ECG system used to capture essential cardiac information such as heart rate and basic arrhythmias. While a twelve-lead system offers a more comprehensive view by capturing signals from multiple angles, the three-lead system provides a simpler, more practical solution for continuous remote monitoring and real-time analysis. It reduces system complexity and patient discomfort, making it suitable for long-term use.

### B. Software Design

The software component manages real-time data acquisition, graphical display, and user interaction. Developed using Python, it leverages various libraries to handle data processing, encryption, and user interface functionalities.

*1)*   Real-Time Data Visualization and Recording: The software facilitates continuous monitoring and recording of ECG signals, providing real-time visualization of heart activity and ensuring data integrity. (see Table I for examples of received data). Devel-oped using Python and the Tkinter library, the software features a graphical user interface (GUI) that dynamically displays ECG signals in real-time. (see Table II for ECG wave sample points). The GUI includes intuitive controls, such as start, stop buttons, to manage data acquisition, and provides a real-time plot for visualizing heart activity. The system continuously records incoming ECG data along with precise timestamps for subsequent analysis.

TABLE I. THE ECG WAVE SAMPLE POINTS

| Marker byte | Meaning of following byte(s) |
|---|---|
| 0xF8 | wave sample points follow |
| 0xFA | Pulse value follows |
| 0xFB | Info byte follows |
| 0x11 | The only info byte defined is 0x11, "LEAD OFF" |

TABLE II. EXAMPLE OF A RECEIVED DATA STREAM AT THE HOST SIDE

| | Data Stream | | | | |
|---|---|---|---|---|---|
| **Byte** | 0xF8 | 0x20 0x23 0x25 | 0xFA 0x80 | 0xF8 | 0x24 0x25 0x26 |
| **Mean** | Wave Marker | ECG sample points | Pulse = 128 | Wave Marker | ECG Sample points |
| **Time** | ............................................................................⟩ | | | | |

The dashed arrow in the "Time" row indicates the continuous flow and sequential order of data over time. The arrow represents the continuity of data flow and its alignment with a time axis. For instance, the first "Wave Marker" (0xF8) and "ECG sample points" (0x20 0x23 0x25) received initially, followed by the "Pulse value" (0xFA 0x80) and the subsequent "Wave Marker" and "ECG sample points" (0x24 0x25 0x26) as time progresses.

*a) Bandpass Filtering in ECG Signal Processing:* In the real-time ECG monitoring and analysis system, the use of a bandpass filter eliminates noise from ECG signals [12]. In the referenced work [13], the authors used a high-pass filter to remove high-frequency noise from the ECG signal output by the preamplifier circuit. They designed a fourth-order Butterworth high-pass filter specifically for this purpose, focusing on eliminating unwanted frequencies above a certain threshold. This method is effective for attenuating high-frequency noise, but it may not adequately address low-frequency drifts or baseline wander in the ECG signal.

In contrast, our study employs a bandpass filtering approach, which not only eliminates high-frequency noise but also filters out low-frequency components below the desired range. By using a Butterworth bandpass filter, we ensure a flat frequency response within the passband, preserving the significant features of the ECG signal such as QRS complexes, while removing low-frequency drifts. This approach provides a more comprehensive noise reduction, enhancing the accuracy of heart rate and disease detection compared to a high-pass filter alone. This filter enhances signal quality by removing frequencies outside the desired range, including high-frequency noise and low-frequency drift. By preserving the significant features of the ECG signal, the bandpass filter ensures accurate heart rate and disease detection. The system employs a Butterworth filter for bandpass filtering, selected for its maximally flat frequency response in the passband, which prevents amplitude distortion of the ECG signal [14].

The cut-off frequencies normalized by dividing them by the Nyquist frequency (half of the sampling frequency).

$$Nyquist = 0.5 * fs \qquad (1)$$

$$low = lowcut / Nyquist \qquad (2)$$

$$high = highcut / Nyquist \qquad (3)$$

Using the normalized frequencies, the Butterworth filter coefficients are computed.

$$(b,a) = butter(order, [low, high], btype = 'band')$$

Here, b and a are the filter coefficients. The filter applied to the signal using the filtfilt function, which performs forward and backward filtering to minimize phase distortion.

$$y = filtfilt(b, a, data)$$

The butter function in SciPy calculates the filter coefficients using the normalized cutoff frequencies. The filtfilt function applies the filter to the signal in a zero-phase filtering manner, which helps in preserving the phase characteristics of the ECG signal.

*2) Disease Detection Model:* The disease detection model integrated into the real-time ECG monitoring system provides immediate analysis of ECG signals to identify potential arrhythmias and other heart conditions. This model, trained on the MIT-BIH Arrhythmia Database, employs a convolutional neural network (CNN) for classification [15][16]. The training process involves extracting ECG segments centered around detected R-peaks with a window size of 180 samples. These segments are labeled based on five primary arrhythmia classes: Normal (N), Left bundle branch block (L), Right bundle branch block (R), Atrial premature contraction (A), and Ventricular premature contraction (V). The data normalized and divided into training and testing sets to ensure reliability. We used the wfdb library to read ECG records and annotations, focusing on the primary signal from the first channel. Using the GQRS algorithm, we identified R-peaks and segmented them accordingly. The CNN model, which includes one convolutional layer, dropout, and dense layers, trained for 10 epochs with a batch size of 32 and evaluated using a validation set to monitor performance and prevent overfitting. For real-time monitoring, we continuously read ECG signals via a serial port, updating the display and predicting arrhythmias using the trained CNN model. This configuration enables precise real-time classification and effective disease detection. We chose Convolutional Neural Networks (CNNs) for ECG signal classification because they efficiently capture spatial features and patterns within the data. CNNs excel at analyzing sequential data like ECG signals, as they can automatically learn to identify essential features such as QRS complexes, P waves, and T waves, which are crucial for detecting various cardiac conditions. Unlike LSTMs, which focus on temporal dependencies and often have longer training and inference times, CNNs offer a good balance between accuracy and computational efficiency. Their ability to process data quickly and accurately makes them ideal for real-time applications, where rapid and reliable analysis is crucial for timely medical decisions. By using CNNs, we ensure that the system can deliver quick and precise predictions making it a suitable choice for real-time tracking.

*3) Privacy-Preserving Health Monitoring:* Remote patient monitoring has become increasingly important in modern healthcare, providing continuous observation of patients' health data outside traditional clinical settings. Ensuring the privacy and security of this sensitive health data is paramount, especially during transmission over potentially insecure networks. To address these concerns, we conducted two different studies utilizing Standard AES and TLS with ECDH & AES encryption methods to secure the transmission of real-time ECG data from patients to healthcare providers. Standard AES encryption to secure the data, a widely used and robust encryption method known for its efficiency and simplicity. However, it relies on pre-shared keys, which can be vulnerable to interception during transmission. Encrypt ECG data using AES for security, generating a unique key for each session. Store encrypted data in a MySQL database with identifiers and timestamps. Authorized users can decrypt the data for medical analysis. The system includes real-time visualization, plotting both original and decrypted ECG data for continuous monitoring and accurate diagnosis. The process of decrypting data for medical review begins with retrieving the encrypted ECG data from the cloud storage by querying the database for records associated with a specific patient or time period. Figure 2 illustrates the systematic processes for encrypting and decrypting ECG data using AES for medical review.

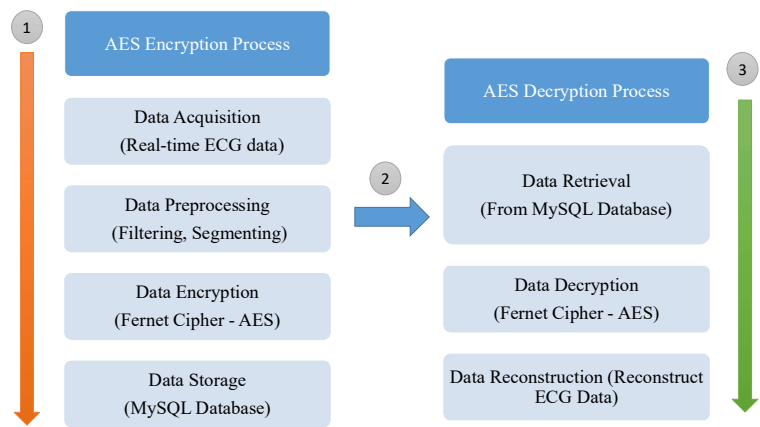

Fig 2.   AES Encryption and Decryption Processes for ECG Data

The second study integrates Elliptic Curve Diffie-Hellman (ECDH) for secure key exchange with AES encryption within a TLS framework, providing an additional layer of security through authenticated and encrypted communication channels. This combination aims to address the key exchange vulnerabilities inherent in the Standard AES approach. The proposed system consists of three main components: the patient device, the server, and the doctor device. The patient device reads ECG data, encrypts it using AES, and transmits the encrypted data to the server over a secure TLS connection. The server stores the encrypted data in a database, and the doctor device retrieves and decrypts the data for analysis. By using TLS for communication, the system ensures that all data transmitted between the patient device, server, and doctor device is encrypted and secure. The ECDH key exchange mechanism provides a strong method for securely exchanging encryption keys, reducing the risk of key interception during transmission. However, integration of ECDH, AES, and TLS adds complexity to the system, the encryption and decryption processes along with the secure key exchange, introduce addi-tional computational overhead. This may impact the performance of the system, especially in resource-constrained environments such as mobile devices. To evaluate the performance and security of the proposed system, we consider various metrics and present the results in Figure 3. These metrics highlight the differences between using standard AES encryption and the proposed system with TLS and ECDH.

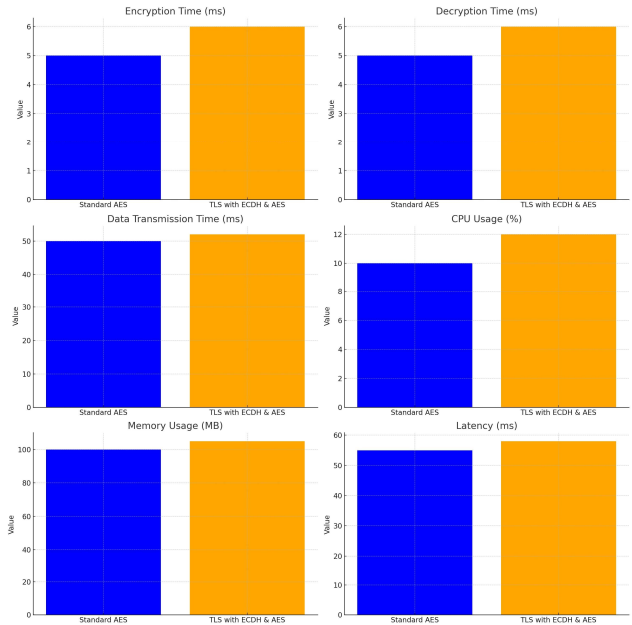

Fig 3. Comparison of Standard AES and TLS with ECDH & AES for Various Metrics

*4)* Homomorphic Encryption for Analysis: The system encrypts ECG data using Fully Homomorphic Encryption (FHE) with the CKKS scheme [17] before storing it in a cloud-based MySQL database. This approach preserves data privacy while allowing computations on encrypted data. Encrypted data is linked to unique IDs and timestamps, ensuring secure, scalable, and efficient storage and retrieval. Authorized medical professionals can decrypt and analyze the data, maintaining confidentiality and enabling advanced analysis capabilities. The Homomorphic Encryption for Analysis component enables secure analysis of ECG data while maintaining patient privacy. Fully Homomorphic Encryption (FHE) supports computations on encrypted data, ensuring the protection of sensitive information during analysis.

- Analysis on Encrypted Data: The system can perform several types of analyses directly on encrypted data, including R-peak detection, statistical analysis, frequency analysis, and heart rate variability (HRV) analysis. Conducting these analyses without decrypting the data preserves its confidentiality.

- R-Peak Detection: The system employs the Pan-Tompkins algorithm to detect R-peaks in the encrypted ECG data. This process involves filtering the data, taking its derivative, squaring the result, and applying a moving window integration.

- Statistical Analysis: The system calculates basic statistical measures such as mean, standard deviation, median, minimum, and maximum values on the encrypted data.

- Frequency Analysis: The system analyzes the frequency components of the encrypted data using the Fourier Transform, identifying the predominant frequencies present in the ECG signal.
- Heart Rate Variability (HRV) Analysis: HRV analysis involves detecting R-peaks, calculating RR intervals, and then computing the mean and standard deviation of the RR intervals.

*Computational Overhead and Performance Trade-Offs:* While FHE provides unparalleled data privacy, it comes with a significant computational cost. The complex mathematical operations required for homomorphic encryption, such as polynomial multiplications and additions in a high-dimensional space, increase both processing time and memory usage. This results in higher latency and reduced throughput compared to traditional encryption methods like AES. For instance, conducting statistical analyses on encrypted data, such as mean and standard deviation calculations, can take several orders of magnitude longer than on unencrypted data.

To address these challenges, we have optimized the system by using the CKKS scheme, which is specifically designed for approximate arithmetic on encrypted data, reducing the computational complexity. Additionally, we have leveraged parallel processing and cloud computing resources to distribute the workload, which significantly reduces the processing time. Despite these optimizations, the system's performance remains sensitive to the size of the dataset and the complexity of the analyses. Therefore, a balance between data security and computational efficiency must be maintained. In a real-world clinical setting, the feasibility of using FHE for continuous real-time monitoring may be limited by the available computational resources and network bandwidth.

## IV. Experimental Results

In this section, we present the experimental setup, data processing pipeline, and results obtained from the real-time ECG data acquisition system. The program has three different interfaces for three different functions. The system's GUI, shown in Figure 4, allows users to start and stop data acquisition, apply filters, and monitor various performance metrics. The interactive elements include buttons for initiating data reading, checkboxes for applying filters, and labels displaying predicted classes, pulse values, processing latency and processed data count. In the following figure, the remote monitoring system with AES visualized.

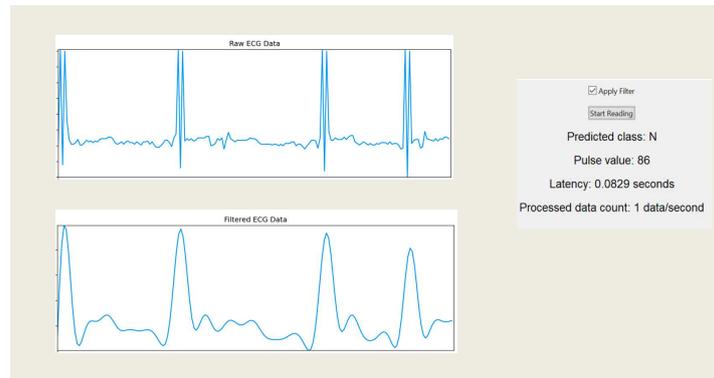

Fig. 4. Raw / Filtered ECG Signals and Real-Time ECG Monitoring System Interface

The Fig. 4 illustrates real-time ECG data acquisition and processing. The top subplot displays the raw ECG signals collected from a three-lead ECG preamplifier connected to a serial port, with data sampled at 50 Hz and presented in their original form. The filtering process helps in isolating the relevant cardiac signals for further analysis.

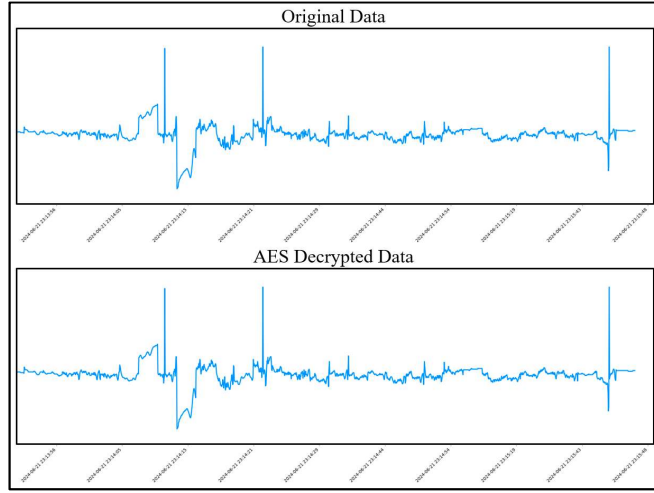

Fig. 5. Original and AES Decrypted ECG Data with Timestamps

The Fig. 5 illustrates The ECG data, sampled at 50 Hz, were filtered, encrypted with AES encryption, and stored in a MySQL database. The system successfully decrypted and plotted the real-time ECG signals on two separate figures: "Original Data" and "AES Decrypted Data," each annotated with timestamp information.

Table III shows the comparison of downstream task results for both original ECG data and encrypted ECG data using FHE. The results indicate a 100% match in R-peak detection, demonstrating that the encryption process preserves critical signal features essential for heart rhythm analysis and arrhythmia identification. Statistical analysis results show a high similarity between original and encrypted data, with a mean value comparison ratio of 99.17%, standard deviation of 98.11%, median of 100%, minimum of 93.75%, and maximum of 94.12%. These findings confirm that the encryption maintains data integrity, ensuring the reliability of analyses conducted on encrypted data. Frequency analysis results indicate a 100% match in dominant frequency components, which is crucial for accurate frequency-based analyses used in diagnosing heart conditions. The HRV analysis also shows a 100% match in both mean and standard deviation of RR intervals, indicating that the encryption process does not affect HRV analysis, thus providing reliable results for clinical decision support systems. These results collectively demonstrate that Fully Homomorphic Encryption (FHE) effectively preserves critical features of ECG data during encryption, enabling accurate and reliable downstream analyses without compromising data integrity or clinical applicability. This confirms FHE as a suitable method for encrypting and analyzing ECG data in a clinical setting, providing both data security and reliable analytical outcomes, which is particularly advantageous in healthcare where data confidentiality is paramount.

Therefore, FHE's implementation in ECG data encryption and analysis proves to be both secure and effective, supporting its use in secure health monitoring systems.

TABLE III. DOWNSTREAM TASK RESULTS

| \multicolumn{4}{c}{Downstream Task Results} ||||
|---|---|---|---|
| *Metric/Analysis* | *Original Data* | *Encrypted Data* | *Comparison Ratio* |
| **R-Peak Detection** | Peaks: [100, 300, 500, ...] | Peaks: [100, 300, 500, ...] | 100% |
| **Statistical Analysis** | | | |
| Mean | 0.0012 | 0.0011 | 99.17 % |
| Standard Deviation | 0.052 | 0.053 | 98.11 % |
| Median | 0.0005 | 0.0005 | 100 % |
| Minimum | -0.15 | -0.16 | 93.75 % |
| Maximum | 0.17 | 0.16 | 94.12 % |
| **Frequency Analysis** | | | |
| Dominant Frequencies (Hz) | 1 Hz, 3 Hz, 5 Hz | 1 Hz, 3 Hz, 5 Hz | 100 % |
| **HRV Analysis** | | | |
| Mean RR Interval (seconds) | 0.85 | 0.85 | 100 % |
| Std Dev RR Interval (seconds) | 0.05 | 0.05 | 100 % |

## V. Conclusions and Limitations

In our real-time ECG signal system, we leverage the MIT-BIH Arrhythmia Database and CNN architectures for accurate arrhythmia classification. The system converts byte streams from the serial port into normalized segments for real-time monitoring, ensuring both accuracy and privacy. Key contributions include a three-lead ECG preamplifier with a serial port for continuous data collection, real-time secured visualization, and data integrity. AES encryption plays a crucial role in this system by securing data before transmission, ensuring that sensitive patient information remains protected during transfer. Using Elliptic Curve Diffie-Hellman (ECDH) for secure key exchange within a TLS framework, the system provides secured encryption and decryption processes, further safeguarding the data integrity and privacy during communication. This ensures that even during transmission, the data protected from unauthorized access. Fully Homomorphic Encryption (FHE) ensures patient data confidentiality, storing encrypted data securely in a cloud-based MySQL database. The CNN model, trained on the MIT-BIH Arrhythmia Database, accurately diagnoses potential arrhythmias, facilitating early intervention and safeguarding patient privacy. The combined use of AES for real-time data encryption and FHE for storage encryption ensures a comprehensive security approach, maintaining both data privacy and integrity throughout the system.

The proposed system, while effective in real-time ECG monitoring and secure data analysis, has several limitations. The use of a three-lead configuration, as opposed to a twelve-lead system, restricts its ability to capture comprehensive cardiac information. The implementation of Fully Homomorphic Encryption (FHE), though crucial for privacy, introduces significant computational overhead, potentially affecting real-time performance, especially in resource-limited environments. The system's reliance on stable network connections for data transmission and its dependence on external databases like the MIT-BIH Arrhythmia Database for model accuracy further limit its scalability and generalizability. Additionally, issues such as battery consumption for mobile applications, the complexity of encryption key management, and potential signal qual-ity loss due to factors like electrode placement and noise may impact the system's robustness and reliability in diverse clinical settings. These challenges necessitate further optimization and validation to enhance the system's efficiency, scalability, and applicability in real-world healthcare scenarios.

## VI. Future Works

Future improvements can focus on expanding the training dataset to include diverse arrhythmias and patient demographics, enhancing the model's generalizability. Federated Learning with larger, varied databases can ensure privacy-preserving training. Integrating with wearable ECG devices for continuous monitoring requires developing wireless communication and robust encryption protocols. Advanced signal-processing techniques like adaptive filtering and machine learning-based noise reduction can improve ECG signal quality and disease detection accuracy. Clinical trials are necessary to validate system performance in real-world settings. Enhancing the graphical user interface (GUI) with intuitive controls, trend analysis, and automatic alerts will improve user experience. Finally, increasing system scalability and integrating sophisticated cloud computing solutions will support larger healthcare networks and real-time analysis on a larger scale.

## Acknowledgment

The system's source code is available as open-source, encouraging further research and development in the field of secure healthcare systems. https://github.com/bbyuksel/ECG-PPS